\documentclass[runningheads]{llncs}
\usepackage[a4paper, total={6.5in, 8in}]{geometry}
\usepackage{graphicx}
\usepackage[colorlinks=true,linkcolor=blue]{hyperref}
\usepackage{amsmath}
\begin{document}
\title{User Logic Development for the Muon Identifier Common Readout Unit for the ALICE Experiment at the Large Hadron Collider\thanks{Supported by NRF-iThemba LABS and SA-CERN Programme}}
%
%
\pagestyle{plain}

\author{Nathan Boyles\inst{1,2} \and
Zinhle Buthelezi\inst{2} \and
Simon Winberg\inst{1} \and
Amit Mishra\inst{1}}
%
\institute{University of Cape Town, Rondebosch, Cape Town, 7700, South Africa 
\url{https://www.uct.ac.za} \and
iThemba LABS, Old Faure Rd, Faure, Cape Town, 7100, South Africa
\url{http://www.tlabs.ac.za}}
\maketitle              
\begin{abstract}
The Large Hadron Collider (LHC) at CERN is undergoing a major upgrade with the goal of increasing the luminosity as more statistics are needed for precision measurements. The presented work pertains to the corresponding upgrade of the ALICE Muon Trigger (MTR) Detector, now named the Muon Identifier (MID).  Previously operated in a triggered readout manner, this detector has transitioned to continuous readout with time-delimited data payloads. However, this results in data rates much higher than the previous operation and hence a new Online-Offline (O2) computing system is also being developed for real-time data processing to reduce the storage requirements.
\newline \newline
Part of the O2 System is based on FPGA technology and is known as the Common Readout Unit (CRU). Being common to many detectors necessitates the development of custom user logic per detector. This work concerns the development of the ALICE MID user logic which will interface to the core CRU firmware and perform the required data processing. It presents the development of a conceptual design and a prototype for the user logic. The resulting prototype shows the ability to meet the established requirements in an effective and optimized manner. Additionally, the modular design approach employed, allows for more features to be easily introduced.


\end{abstract}

\section{Introduction and Background}

ALICE -A Large Ion Collider Experiment - is one of the four major experiments at the LHC at CERN. Its purpose is to advance our understanding of the universe by colliding ions together and studying their constituents. The work described in this report pertains to the development of the user logic of the ALICE MID CRU \cite{upali}.  This logic is a custom component specific to the Muon Identifier to perform data processing functions and forms part of the upgrades to the ALICE detector. While some systems will be replaced, most other detectors, including the Muon Spectrometer (Tracking and Trigger systems) will receive new front-end readout electronics\cite{psc}.
\newline \newline 
The ALICE detector has historically operated in a triggered manner when physics events of interest are detected at the front-end electronics (FEE). This method of data acquisition - using physics triggers, will become very inefficient in view of the upgraded ALICE detector due to the rate of data production. The upgrade strategy is based on collecting Pb-Pb collisions at luminosities up to 6x10$^{27}$cm$^{-2}$s$^{-1}$ , corresponding to collision rates of 50 kHz, where collision data will be transferred to  the  O2 system either upon a minimum bias trigger or in a self-triggered, continuous way\cite{upali}. The upgraded detectors will be self-triggered and operate in a continuous readout mode making use of heartbeat triggers at fixed intervals in time \cite{upali}.  Furthermore, the entire data acquisition chain has to be improved to accommodate this new readout method. Most notably, a new initial real-time data-processing unit based on FPGA technology called the CRU will be implemented. The CRU will be common to all detectors and is the central interface between the Front-End System (FES), Central Trigger System (CTS) and the O2 system.

\subsection{The Muon Trigger System}
The muon trigger (MTR) system is a part of the muon spectrometer, the physics aim of which is to study single and di-muon events produced in the decay of various subatomic particles.  The MTR was designed for muon identification and to select single and di-muons with transverse momentum above a programmable threshold. The values of the programmable momentum thresholds can vary between 0.5 to 4 GeV/c. The ALICE Central Trigger Processor (CTP) receives six trigger signals less than 800 ns after the interaction, at 40 MHz \cite{ali}.

\begin{figure}[h]
\centering
\includegraphics[width=0.4\textwidth]{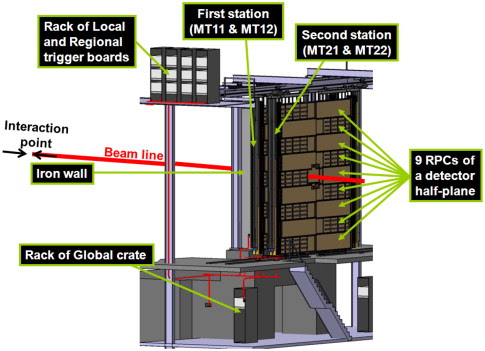}
\caption{ALICE MTR System \cite{rpc_fig} }
\label{fig:The MTR detector}
\end{figure}

The MTR (shown in Fig. \ref{fig:The MTR detector}) consists of four planes of Resistive Plate Chambers (RPC) arranged in two stations one metre apart. A single plane comprised of 18 RPC's typically 70x300 cm$^2$ in size, resulting in a total active area of 140 m$^2$. Momentum selection is performed using a position-sensitive trigger detector with spatial resolution less than 1 cm. Cartesian coordinates of RPC hits are determined using segmented strips with pitch and length increasing with their distance from the beam axis.
\newline \newline
The RPC's are an air-tight parallel-plate detector with a single gap of 2 mm. The plates are made of low resistivity bakelite (${\sim}10^9\Omega$ cm)  and are filled with a gas mixture composed of  89.7\% $C_2H_2F_4$, 10\% $C_4H_{10}$ and 0.3\% $SF_6$. The gas relative humidity is kept at 37 \% to prevent variations in the bakelite resistivity. The high operating voltages are optimized for each chamber and range from 10.0 to 10.4 kV \cite{rpc}. The RPC's are equipped with dual-threshold front-end discriminators which have been tailored to the timing properties of the detector and reach the necessary time resolution (1--2 ns) for the identification of the bunch crossing (BC). 
\newline \newline
The discriminators then send signals to the trigger electronics working in a pipeline mode at 40 MHz. The trigger electronics i.e. the local, regional and global boards are arranged hierarchically merging data along the datapath. The trigger algorithm searches for a single muon track using the information of the four RPC's planes that approximately originates from the primary interaction vertex. Additionally, hits in at least 3 of the 4 detector planes in both the bending and non-bending (referring to the axis along which the muons bend due to the presence of the magnetic field in the muon spectrometer) plane are required to define a track. The regional and global levels gather and deliver all the signals from the local boards of the entire detector.
 
\section{Methodology}
\label{sec:meth}

The present work used the results of earlier work on periphery systems which set the user logic constraints such as the FPGA, communication protocol, resource utilization etc. The knowledge acquired from the systematic studies on these periphery systems facilitated a seamless integration into the MID System. By reviewing technical design reports, technical notes, presentations as well as consulting hardware and software developers working on the complimentary parts of the data acquisition chain, a systems engineering approach was used to converge on an appropriate user logic design.

\subsection{User Requirements}
The data payload is delivered to the core CRU firmware (FW) via 16 optical links at 3.2Gb/s using the standard Gigabit Transceiver (GBT) protocol \cite{gbtxman}. Subsequently, the data is deserialized and transmitted to the user logic along a bus. 

This data contains the header information from the specific segment of the 4 planes connected to that specific card such as status, trigger, timing and card ID information as well as the digitized muon hits in the case of a local card \cite{midro}. The card ID indicates the regional or local card ID from which the data payload originates. However, in the case of the local card this ID is relative to its associated regional card and therefore not unique globally. The user logic is required to determine a unique ID using the regional and local ID.
\newline \newline 
Slight variations in the properties of the multiple optical and electrical links introduces variations in transmission time. Consequently, the data is out of sync on arrival at the CRU and needs to be synchronized in the user logic.
\newline \newline
In continuous readout mode data acquired is time delimited using a periodic heartbeat (HB) trigger transmitted by the Central Trigger System every ${\sim}$89us \cite{fcost}. The period between heartbeat triggers is known as a heartbeat frame (HBF). Upon receiving the start of continuous (SOC) trigger, the FEE initiatess the digitization of its channels periodically but at a higher frequency than that on the HB triggers. The data being transmitted contains this internal bunch count of the FEE but not the HB count. This effectively produces data payload in sub-heartbeat frames. This is incompatible with the O2 system since it requires heartbeat IDs (HBID) in the Raw Data Header (RDH). Thus, the HBID needs to be determined using the internal bunch counter information provided in the FEE data payload as well as Bunch Crossing ID (BCID) in the CTS payload. Furthermore, the RDH must be populated when an orbit trigger is received, or when a page is more than 8 kB, in which case a new page is opened with an RDH on top of it which is a copy of the previous one with the page counter incremented.

The trigger that produced that set of data also needs to be added to the RDH which is a simple matter of populating the respective field.
\newline \newline
A key specification on the core CRU FW PCIe downlink is the packetization of the data payload (limited to 8kB) prior to transmission.

\subsection{Design Specification}

Since data is streaming continuously, the user logic must monitor the data bus for a valid status byte to initiate data sampling at which time subsequent data bytes will be temporarily stored in registers for processing.  
\newline \newline
Once the Local and Regional Card IDs for that data set have been extracted a valid bit is asserted initiating the start of data reformatting. This results in four O2 headers for the strip patterns corresponding to the four planes of the MID.
\newline \newline
The value stored in the internal bunch counter register is monitored by logic for resets indicating the start of a new HBF and hence an increment in the HBID.
\newline \newline
To achieve synchronization a 2-Dimensional FIFO will be utilized, where an individual FIFO is assigned to each local card. The payload will be temporarily stored here until each FIFO has at least one dataset at which point synchronized readout will begin.
\newline \newline
Synchronic data will be read out using a merge-sort algorithm and will be streamed as an 8kB packet in the format required by the O2 system, consisting of an RDH followed by the payload where data are ordered by an internal bunch counter.

\subsection{Functional Verification}
\begin{figure}[h!]
\centering
\includegraphics[width=0.45\textwidth]{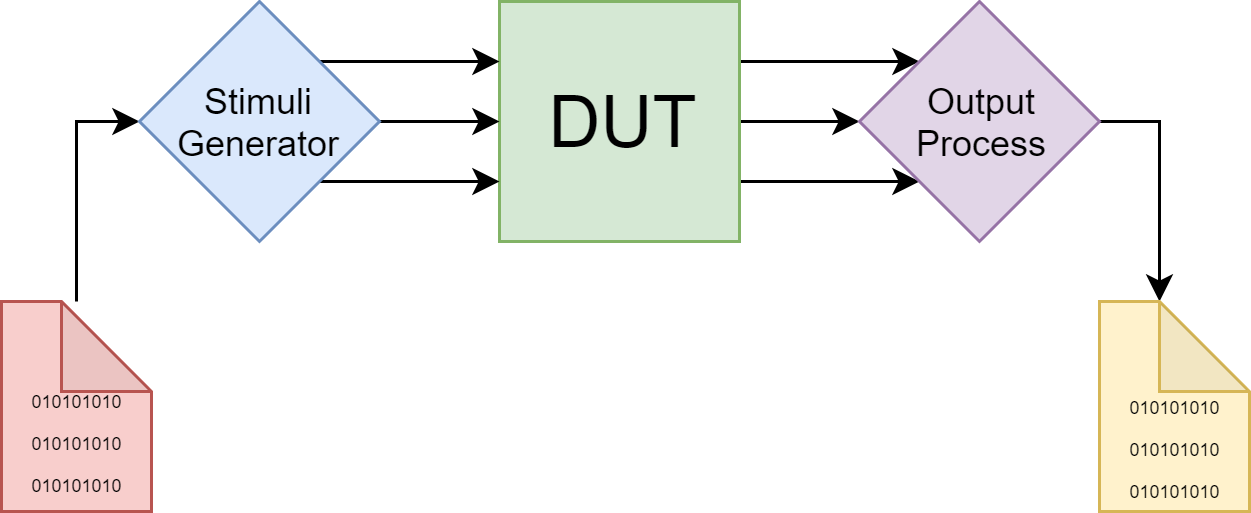}
\caption{Functional verification in HDL testbench}
\label{fig:ver_meth}
\end{figure}

For this work, functional verification is simulation-based, driving vectors into each module using a stimuli generator. Conformance to requirements is then assessed by inspection. A graphical depiction of this procedure is shown in Fig. \ref{fig:ver_meth}.

The suite of tests that need to be performed for functional verification are as follows:

\begin{itemize}
    \item Successful extraction and temporary storage of data payload
    \item Accurate reformatting of data payload 
    \item Successful synchronization of data payload
    \item Successful transmission of data payload from user logic to core CRU FW
\end{itemize}

The single operating scenario that would encapsulate this entire verification suite would be that of the frond-end test (FET). In the MID, a FET is performed to identify dead channels (faulty electronics or broken links) in the data acquisition chain. This is performed periodically by the CTS sending a calibration trigger to the front-end electronics who upon reception of this trigger, fire (transmit a "TRUE" signal) which results in a complete data payload. As such, the FET expected dataset with no dead channels is used as the simulated dataset for functional verification. With the exception of the data extraction and synchronization module, the vectors are always synchronized when driven into module ports. 

\section{System Architecture}
This section outlines a conceptual design design for the MID-CRU user logic.
\subsection{Data Extraction}
\label{sec:dat_extr}
Data extraction is initiated when a valid status byte is detected. This is achieved by monitoring module ports for a valid status byte and sampling the ports every clock cycle when this is true. The data is then stored in registers for processing. 

\subsection{Data Reformatting}
\label{sec:dat_rfmt}

\begin{figure}[h!]
\centering
\includegraphics[width=0.8\textwidth]{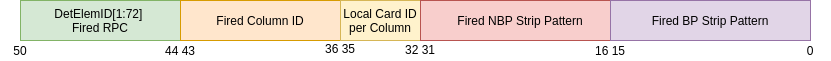}
\caption{O2 Header Format Required \cite{o2_h_c}}
\label{fig:o2_header}
\end{figure}

\begin{figure}[h!]
\centering
\includegraphics[width=0.6\textwidth]{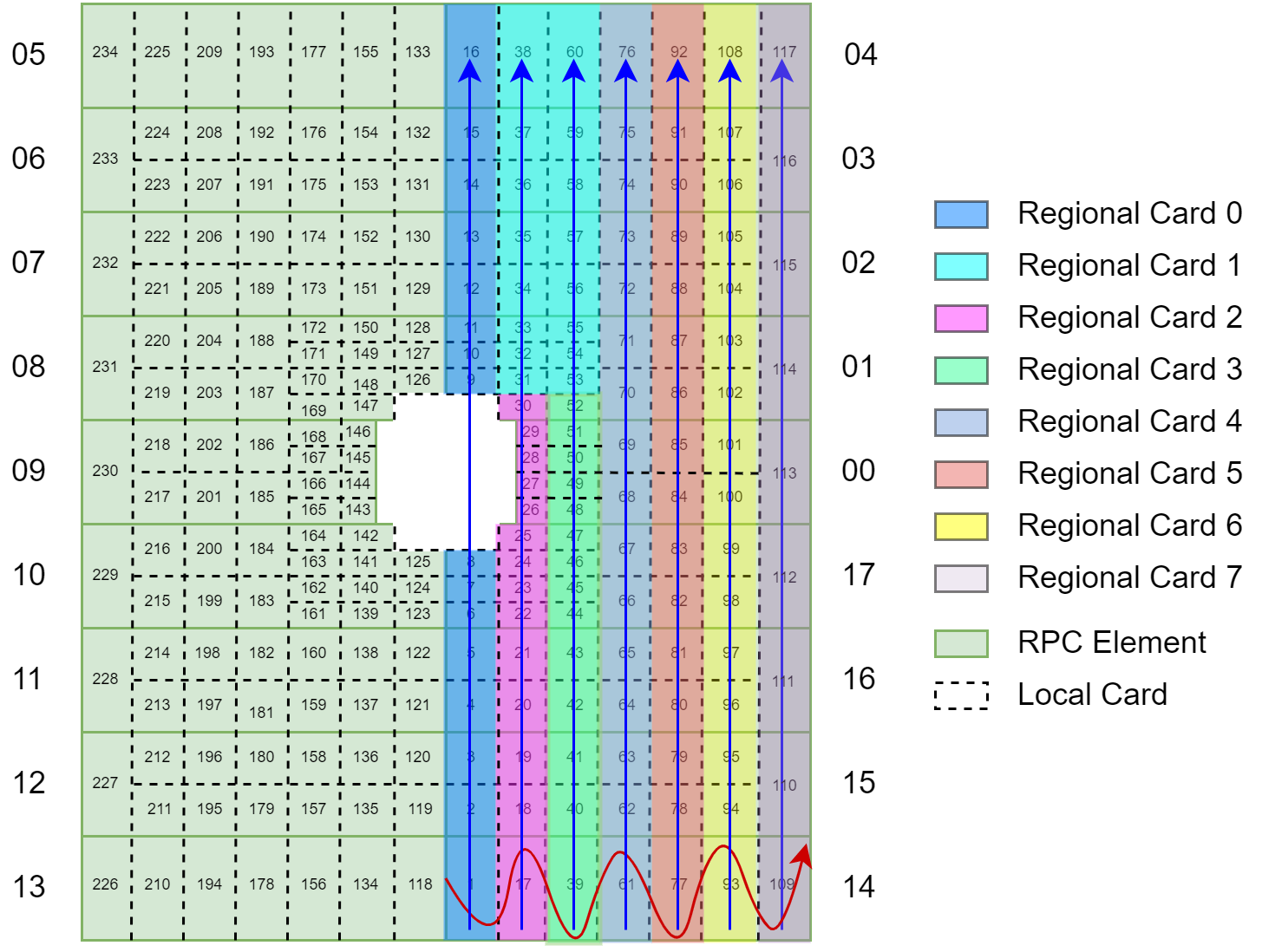}
\caption{Required data readout for O2 Computing System \cite{dat_ro}}
\label{fig:rpc_lo_ro}
\end{figure}

The user logic is required to reformat the data payload received from the core CRU FW to correspond to a unique location in the ALICE MID where the data originates. The FEE protocol contained in the 80 bit data bus - per GBT link - contains byte fragments of the data payload from local card 0 through 7 as well as byte fragments from the 2 internal regional card links. The information necessary to successfully identify a unique detector location is the card ID embedded in the relevant data streams. The required O2 header per dataset can be seen in Fig. \ref{fig:o2_header}. The strip pattern for the bending and non-bending is fully described by indicating the RPC detector element where the data originates (i.e. data per plane is split), the column in that particular RPC and finally the position of the local card within that column.
\newline \newline
When data extraction of the local and regional card ID’s is complete the corresponding
registers where the data is stored is applied to the reformatting component of the user logic and a valid signal
is asserted to start the process. Combinations of regional and local card ID’s are then mapped to four RPC
elements and also the cards location in that RPC element based on detector geometry, the segmentation of
plane MT11 of the MID is shown in Fig. \ref{fig:rpc_lo_ro}.
After successful synthesis of the O2 header each bending plane (BP) and non-bending (NBP) strip
patterns are attached, which were stored in registers during the data extraction process. The reformatted
data is then transmitted to the 2-D FIFO for data synchronization via the reformatting module’s
downlink ports.

\subsection{Zero Suppression}
\label{sec:zspr}

In continuous readout mode, data is being streamed through the data acquisition chain regardless of whether the data is valid and therefore should only initiate sampling when a valid status byte is detected. Additionally, not all local cards are transmitting valid data in normal operation, hence the data stream should not be sampled. Tracklet information in the FEE header indicates which planes are sending data and this information can be used to mask the part of logic not expected to receive data, inject dummy data into the associated FIFO and populate a detector map to reject that data in the final readout. It's anticipated that faulty or broken links can also introduce zero's and hence need to be suppressed. A proposed addition to the core CRU FW is the introduction of input signals to the user logic indicating when a link is broken or faulty. These input signals can be used in the same manner as explained above, namely, masking logic for that GBT block, inserting dummy data and rejecting the dummy data in the final readout.

\subsection{Data Synchronization}
\label{sec:dat_sync}

\begin{figure}[h!]
\centering
\includegraphics[width=0.7\textwidth]{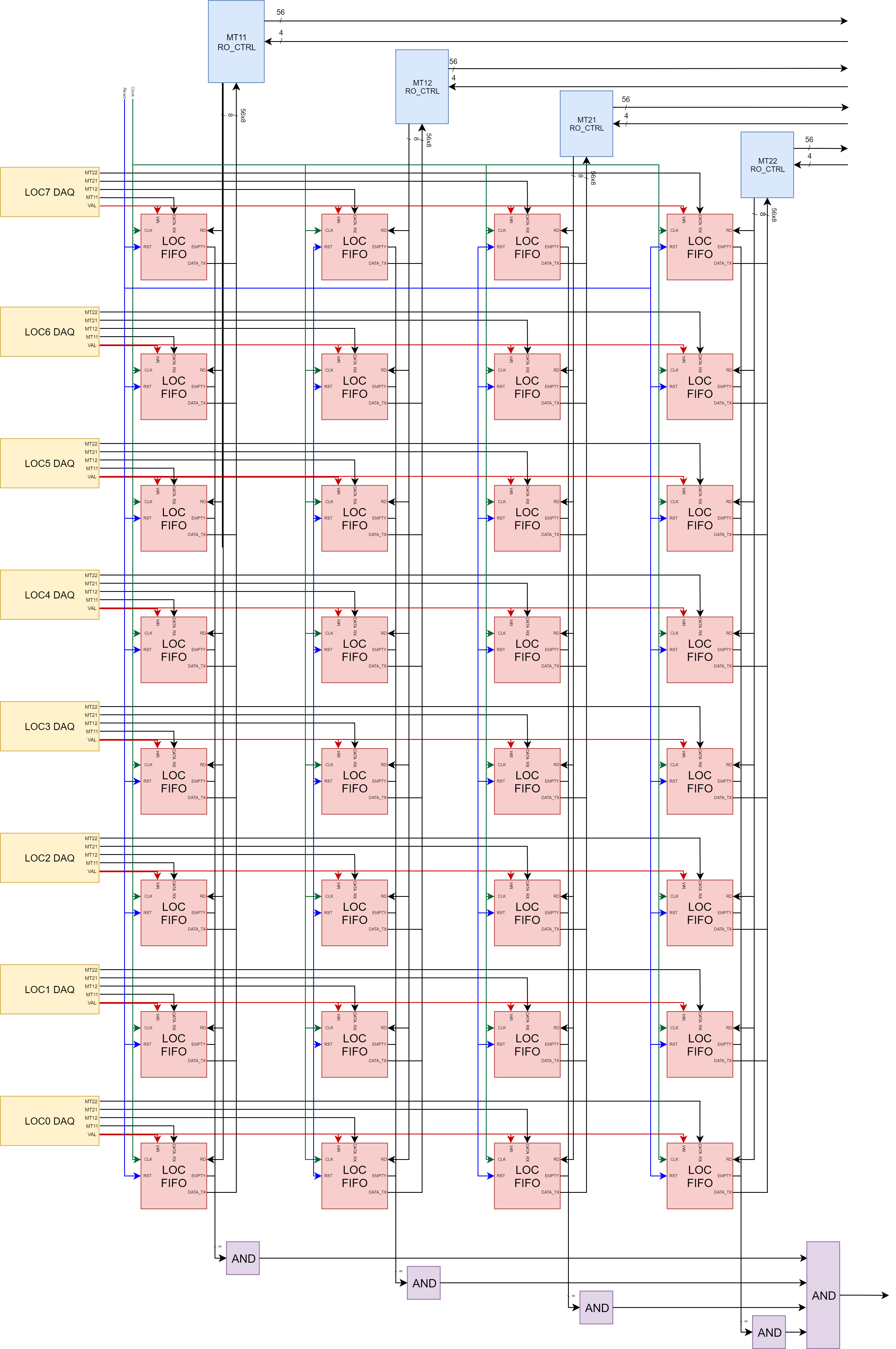}
\caption{Schematic of data synchronization module of MID-CRU User Logic designed in this work}
\label{fig:ro_s1}
\end{figure}

Data de-synchronization occurs due to variations of data transmission in the optical links. To perform data synchronization a 2-D FIFO is used by building an array of FIFO's, where a single FIFO corresponds to data per local card per plane. An implementation of the 2D-FIFO for a single GBT link is shown in Fig. \ref{fig:ro_s1}. The empty signal from each FIFO is used to determine when data has arrived from all local cards at the beginning of data taking and making use of a local AND operation for each local card from a single GBT link and global AND operation on all GBT links. When the global AND is 'TRUE', synchronization is achieved and readout for consecutive internal bunch count can be performed moving each FIFO up one level and repeating this process.

\subsection{Population of Raw Data Header}
\label{sec:p_rdh}

There are fields that need to be populated with information pertaining to the packet of data being transmitted to the core CRU FW, in particular, the heartbeat trigger that produced that data and the heartbeat ID of that particular trigger. There are 2$^{16}$ consecutive datasets within one heartbeat frame with no HBID associated with it. By monitoring resets, in the internal bunch counter, the heartbeat can be determined and populated in the RDH along with the trigger of the data payload. 

\subsection{Data Readout}
\label{sec:data_ro}

To minimise memory requirements as well as latency, an efficient readout mechanism is required. This is achieved using a hardware implementation of a merge-sort algorithm.

\subsubsection{Stage 1 Readout:}
\label{sec:s1}
\begin{figure}[h!]
\centering
\includegraphics[width=0.2\textwidth]{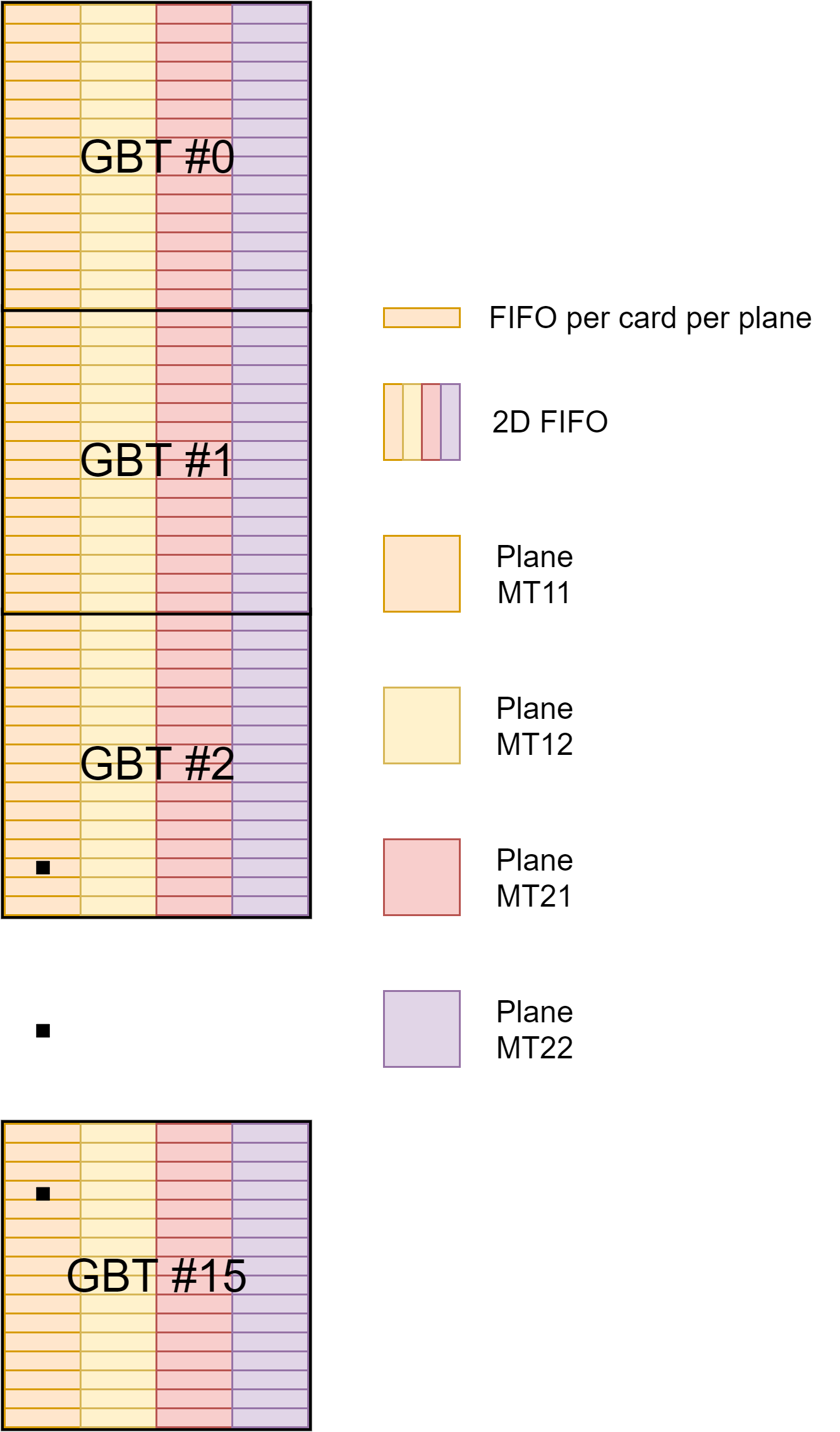}
\caption{Stage 1 Memory Partitioning designed from this work}
\label{fig:pipe}
\end{figure}

Upon data synchronization in the 2-D FIFO, Stage 1 read out is initiated. This is performed in an efficient manner by pipe lining since the data can be partitioned in columns. The segmentation of Stage 1 memory is shown in Fig. \ref{fig:pipe}.

Applying row and column partitioning to the problem allows 2 levels of pipelining. 

Every GBT 2D-FIFO bank can be readout concurrently, additionally all 4 planes per GBT 2D-FIFO bank can also be readout simultaneously. Since there are only 8 local card data sets per plane per GBT 2D-FIFO it would take only 8 clock cycles to readout the Stage 1 memory instead of 8192 clock cycles which is a significant improvement to the throughput.    

\begin{figure}[h!]
\centering
\includegraphics[width=0.45\textwidth]{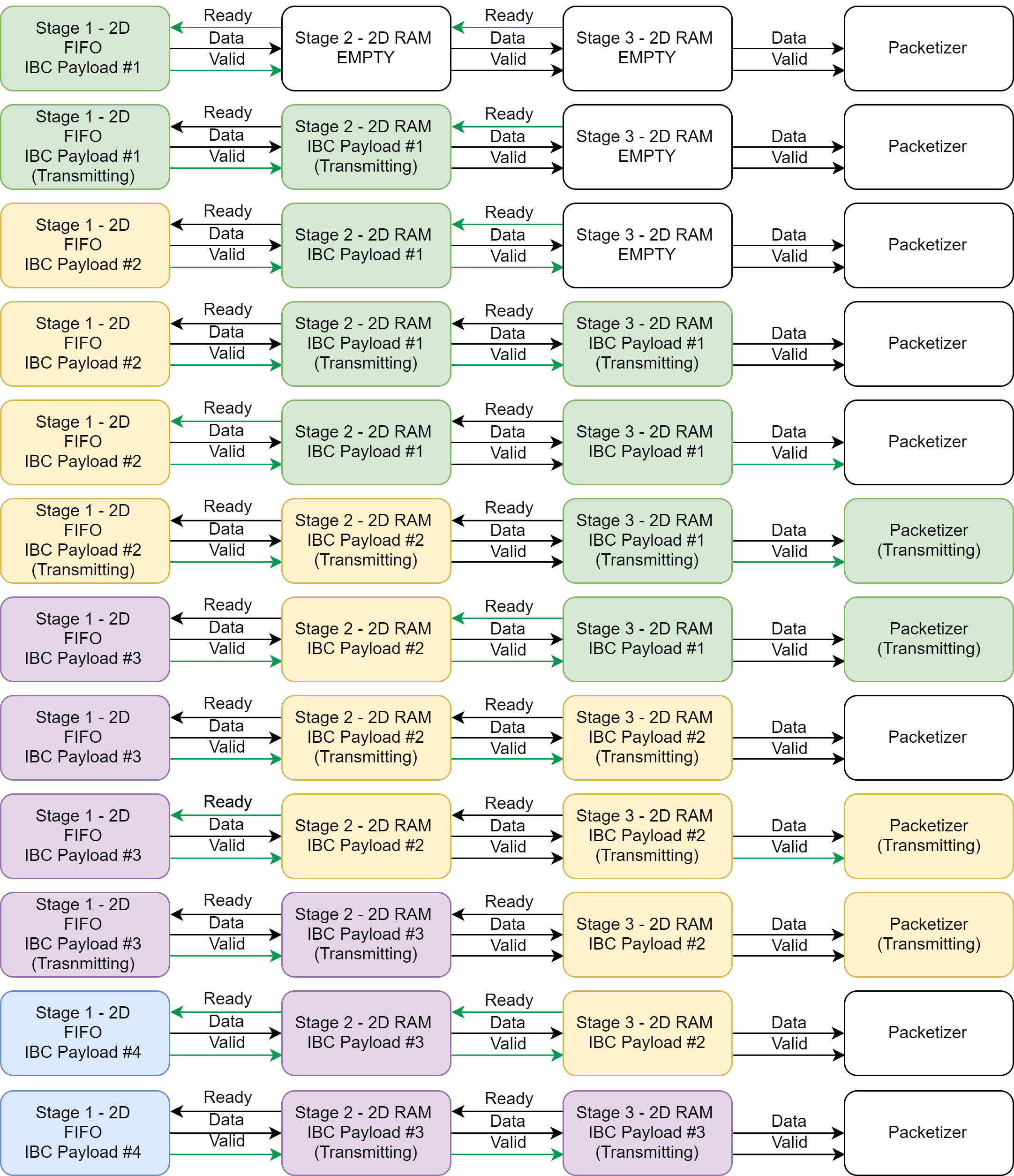}
\caption{Buffered handshaking designed from this work}
\label{fig:ro}
\end{figure}

The readout algorithm also makes use of handshaking signals between stages. This allows constant communication between the stages indicating when they should transmit data to downstream modules or discard data from upstream modules. This process is illustrated in Fig. \ref{fig:ro}. 

\subsubsection{Stage 2 Readout:}
\label{sec:s2}
\begin{figure}[h!]
\centering
\includegraphics[width=0.2\textwidth]{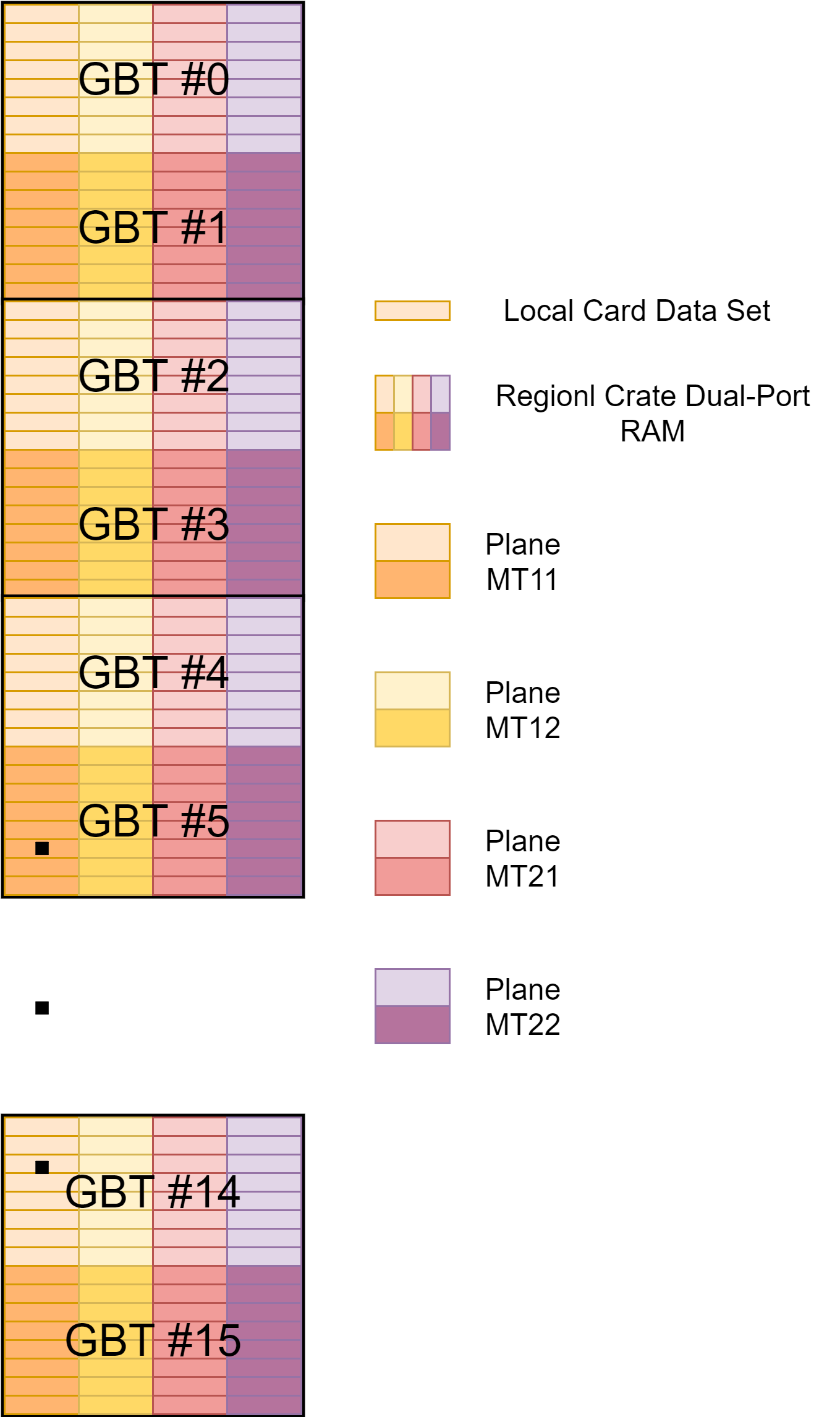}
\caption{Stage 2 Memory Partitioning designed from this work}
\label{fig:pipe2}
\end{figure}

Since 2 regional links correspond to the same regional crate i.e. local cards 0 through 15 dual-port RAM is an obvious choice as the base module for Stage 2 memory. This enables the merging of these data sets from the local cards per plane without the need for two separate RAM modules. The advantage of which is optimized resource utilization as well avoiding additional buffering introduced by another RAM module. The segmentation of the Stage 2 memory is illustrated in Fig. \ref{fig:pipe2}. 

Stage 2 readout initiates when Stage 1 asserts the input valid handshaking signal  (s12\_data\_v\_i) this in turn results in Stage 2 de-asserting the ready for input handshaking signal  (s2\_s3\_busy\_o). 

Regardless of s2\_s3\_busy\_o being deasserted, Stage 1 data readout remains valid and continues until complete, followed by s12\_data\_v\_i being de-asserted. Stage 1 will reassert the input valid signal once the data payload transmitted to Stage 2 memory has been successfully transmitted to Stage 3 memory at which point s2\_s3\_busy\_o is reasserted restarting Stage 1 - Stage 2 readout for the next data payload. This procedure is depicted in Fig. \ref{fig:ro}.

\subsubsection{Stage 3 Readout:}
\begin{figure}[h!]
\centering
\includegraphics[width=0.45\textwidth]{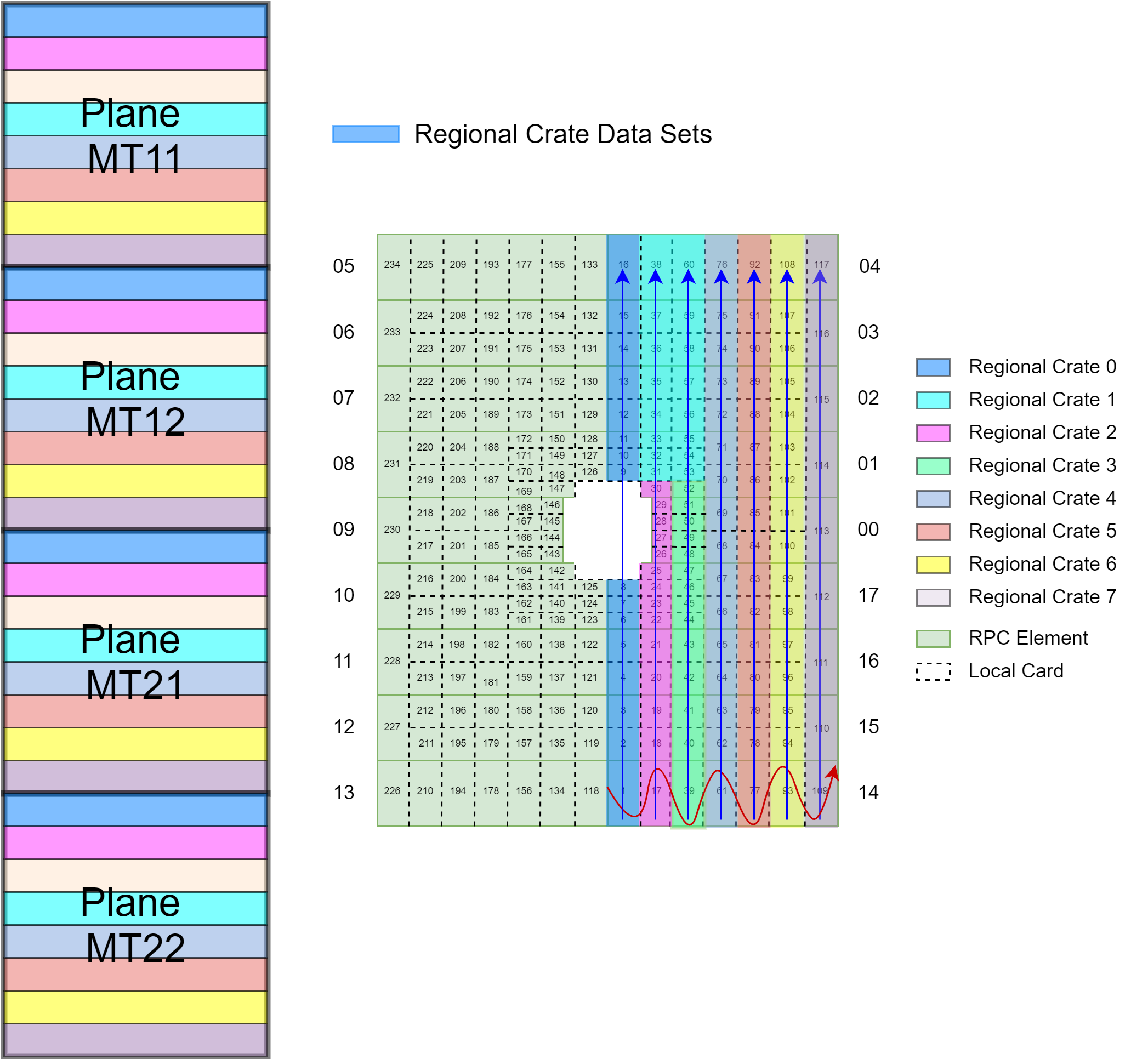}
\caption{Stage 3 Memory Partitioning}
\label{fig:pipe3}
\end{figure}

Stage 3 readout is responsible for merging the data payload from the 8 regional crates from the same plane (MT11, MT12, MT21, MT22) into a single block of memory, depicted in Fig. \ref{fig:pipe3}. The 4 data output signals connect to the 4 data input signals of Stage 3 as well as 2 handshaking signals. 

\newpage
Similarly to Stage 2, Stage 3 readout initiates when it's input valid signal (s23\_data\_v\_i) is asserted by Stage 2 while it's ready for input signal (s3\_busy\_o) is asserted. The 4 columns of the Stage 2 memory are readout concurrently storing the data sets of all the local cards of each regional crate in ascending order in 4 separate RAM modules. Once this process is complete s23\_data\_v\_i is deasserted, s3\_ram\_data\_v\_o asserted and the contents of the 4 RAM modules are readout consecutively in order of plane. When Stage 3 readout is complete s23\_data\_v\_i is asserted, s3\_busy\_o and s3\_ram\_data\_v\_o both deasserted indicating that Stage 3 is ready for the next payload. This process is depicted in Fig. \ref{fig:ro}.  
\subsection{Full Configuration}

\begin{figure}[h!]
\centering
\includegraphics[width=0.7\textwidth]{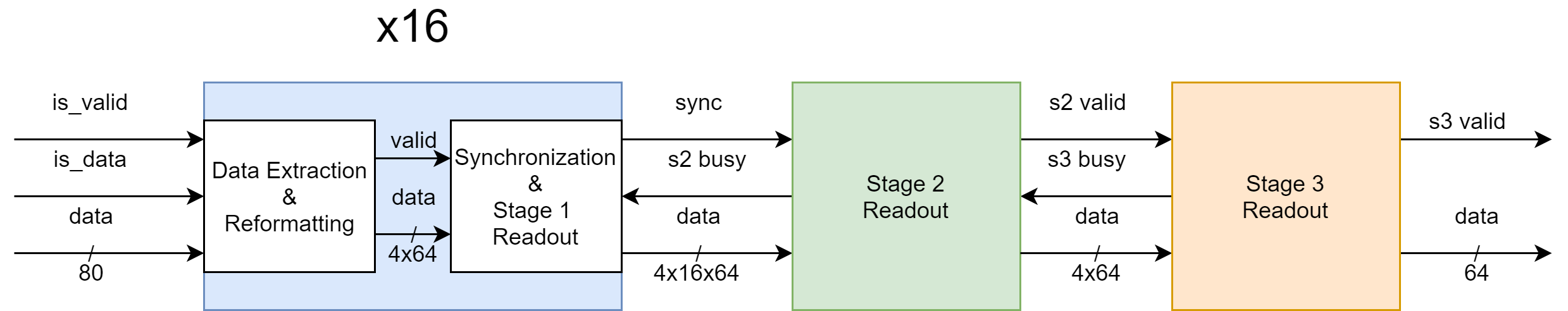}
\caption{Full Configuration}
\label{fig:mid_proto}
\end{figure}

The modules outlined above form base components for the designed MID user logic. A simplified depiction of how they are configured is shown in Fig. \ref{fig:mid_proto}. Each reformatting and synchronization module corresponds to a single GBT link leading to 16 blocks of these. The Stage 2 readout module merges and sorts the data payload for a single internal bunch count according to the regional crate the data came from. Finally Stage 3 readout merges and sorts Stage 2 data according to the plane that the data came from and reads it out on a single bus consecutively.

\section{Results}
This section outlines the results obtained from the system verification procedure outlined previously. It begins with an evaluation of the data extraction system, followed by data reformatting, data synchronization and concludes with an evaluation of the data readout system.
\subsection{Evaluation of Data Extraction}
\label{sec:eval}
\begin{figure}[h!]
\centering
\includegraphics[width=0.7\textwidth]{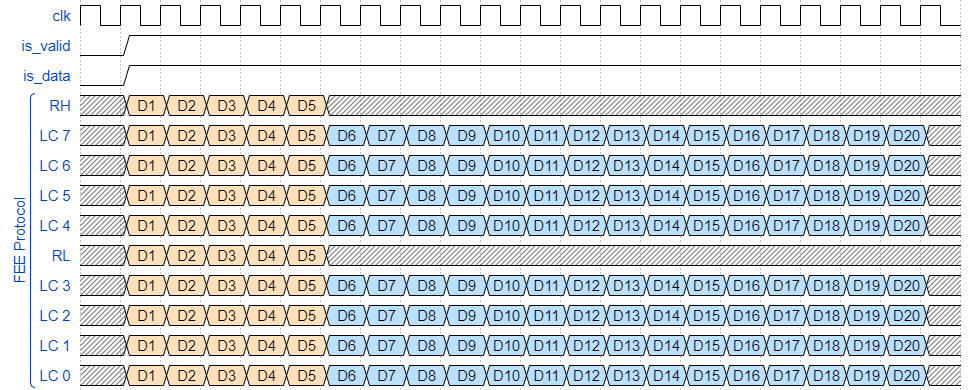}
\caption{Timing diagram depicting test data being driven through data extraction input ports.}
\label{fig:extr_wav}
\end{figure}

The data protocol of the front-end cards is such that the 80 bit field of the GBT link of 10 bytes contains the data payload from the 8 local cards connected to that regional card and two internal links of the card \cite{midro}. On every clock cycle, a byte of the data payload from each card is transmitted to the CRU.
\newline \newline
Using the FEE data format as a reference, expected input data is simulated by reading a text file in the hardware description language (HDL) testbench for this module and driving it into the module via a stimuli generator. The data used can be seen in Fig. \ref{fig:extr_wav}.

Data extraction is achieved by storing these bytes in unique registers for use in other modules of the user logic. In total 9 registers are used per local card. Assuming 128 local cards for simplicity, this totals 1152 registers in the user logic. This module has zero latency since the data is stored in these registers on the same clock it is sampled at the input ports.

\subsection{Evaluation of Data Reformatting}
Data reformatting is achieved in 2 stages, namely, synthesizing the O2 header that uniquely identifies where the data payload originates in the MID detector and attaching the relevant data payload to its respective header.
\newline \newline
After the process of data extraction, the FEE payload has been temporarily stored in the registers. The local and regional card data is driven through a module responsible for ascertaining the unique location where the data originates from lookup tables (LUT's) and providing four O2 headers, one for each plane (see section \ref{sec:dat_rfmt} for more detail). Once the headers are available, their data payloads for each plane are attached to them as they are still being stored in registers. This results in an effective mapping from the FEE data format to the required O2 data format. 
\newline \newline
Testing of the module responsible for synthesizing the O2 header is performed using all possible input combinations of local and regional cards.  

\begin{figure}[h!]
\centering
\includegraphics[width=0.5\textwidth]{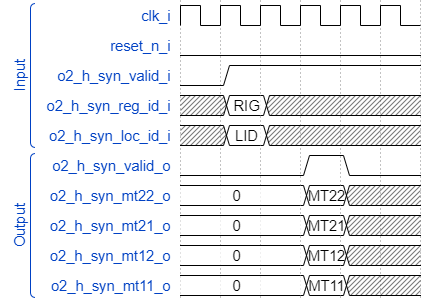}
\caption{Timing diagram depicting test data being driven through O2 header synthesis input ports and output.}
\label{fig:o2_wav}
\end{figure}

The results (shown in Fig. \ref{fig:o2_wav}) show accurate synthesis of a new O2 header and output data of this module produces four sets of data indicating correct operation of the data reformatting module. A latency of 1 clock cycle occurs in this module due to signals updating on positive clock edges. 
\subsection{Evaluation of Data Synchronization}
\begin{figure}[h!]
\centering
\includegraphics[width=0.7\textwidth]{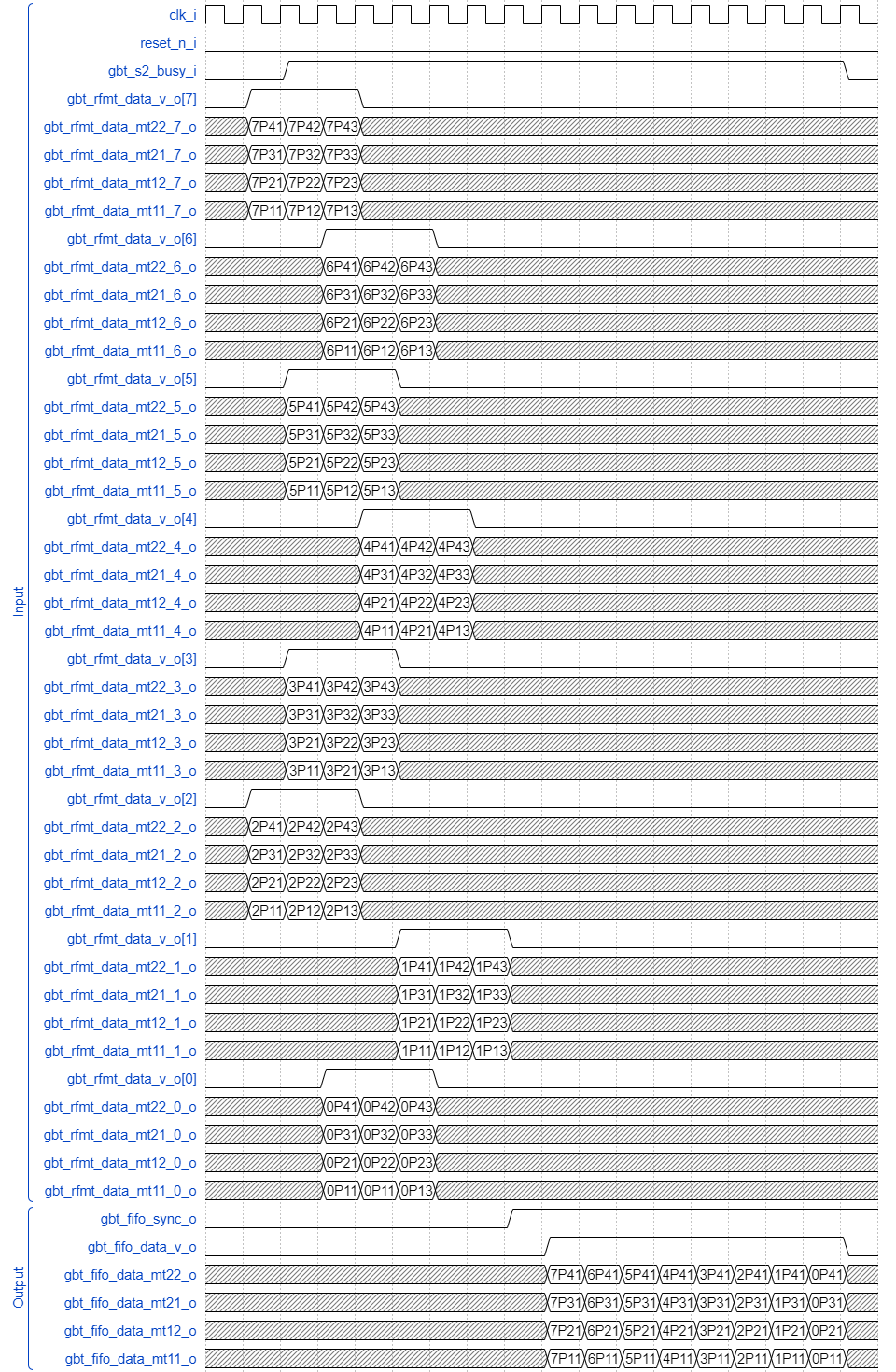}
\caption{Timing diagram depicting test data being driven through synchronization module input ports and output.}
\label{fig:sync_wav}
\end{figure}

In order to evaluate whether data synchronization has been performed correctly in the data synchronization module i.e. the 2D FIFO, data is injected into the ports of this module at varying clock cycles. 
\newline \newline
The newly reformatted data payload, 64 bits in length, for the 8 local cards and 4 planes are read from a text file and driven into the ports of the 2D FIFO module via a stimuli generator. The data payloads for these cards, however, are injected at different times, simulating anticipated transmission delays (explained in section \ref{sec:dat_sync}). This testing procedure can be seen in Fig. \ref{fig:sync_wav}.

Correct data synchronization is evaluated by inspection of output files produced by the HDL testbench of the 2-D FIFO module. The 2-D FIFO is designed to output the data from each plane independently due to column partitioning design methodology and in sync. Moreover, each sub-FIFO per plane is read consecutively on each clock cycle while writing to half the memory of an S2 RAM module with the other half being written to by the 2-D FIFO module corresponding to the second regional card on the same regional crate in parallel.
\newline \newline
The data contained in Fig. \ref{fig:sync_wav} shows this occurring as expected from the 11\textsuperscript{th} clock cycle to the 18\textsuperscript{th} clock cycle.
\newline \newline
The Arria 10 FPGA provides two main memory resources, namely: M20K and MLAB's. Due to the limited amount of M20K memory as well the relatively large number of memory blocks needed for the 2D modules, a logical design decision was to make use of the MLAB memory available. Each 2D FIFO module makes use of 32 memory blocks comprising 4Mb (8 per plane corresponding to 8 local cards). Since there are 16 GBT links per CRU the total amount of memory used is calculated as follows:

\begin{equation}
\resizebox{.9\hsize}{!}{$
    \begin{split}
    Memory\ Used &= (16\ GBT\ links)*(32\ FIFO\ modules)*(64\ words\ depth)*(64\ bits) \\
    &=\ 0.262144 MB
    \end{split}
    $}
\end{equation}

Although not explicitly mentioned, the Arria 10 FPGA provides 1.6 MB of MLAB memory, hence the memory used in for data synchronization i.e. 0.262144 MB accounts for 16.5\% of the MLAB memory resources available. Additionally, a depth of 64 words was chosen arbitrarily during development and 8 words would most likely meet the requirements for synchronization reducing resource usage nearly by a factor of 3 to 5.13\%.
\newline \newline
A rudimentary calculation of latency is performed by calculating the difference between the time at which synchronization (first data payload) occurs and the time at which the final data set from the first clock cycle is transmitted from the 2-D FIFO.
\newline \newline
Figure \ref{fig:sync_wav} shows that the data payload from local card 4 is injected last into the module at the 4\textsuperscript{th} clock cycle and the synchronized data being transmitted at the 18\textsuperscript{th} clock cycle. Hence, there is a latency of 14 clock cycles from input to output i.e. $\sim$ 58ns

\subsection{Evaluation of Readout}

This sections evaluates the data readout mechanism of the user logic.

\subsubsection{Stage 1 Readout:}
Stage 1 readout is performed once synchronization has been achieved in the 2-D FIFO modules. All 16 2-D FIFO modules need to be in sync for this process to start, once that occurs the first word per local card in each plane is read consecutively. The data read from the two 2D FIFO modules corresponding to the same regional crate are written to the top and bottom half of a single dual-port RAM module.
\newline \newline
Output data produced by the 2-D FIFO module is depicted graphically in Fig. \ref{fig:s2_wav}. It can be observed that the data valid signal is asserted whilst valid data is being transmitted from the 11\textsuperscript{th} clock cycle.

\subsubsection{Stage 2 Readout:}

\begin{figure}[h!]
\centering
\includegraphics[width=0.7\textwidth]{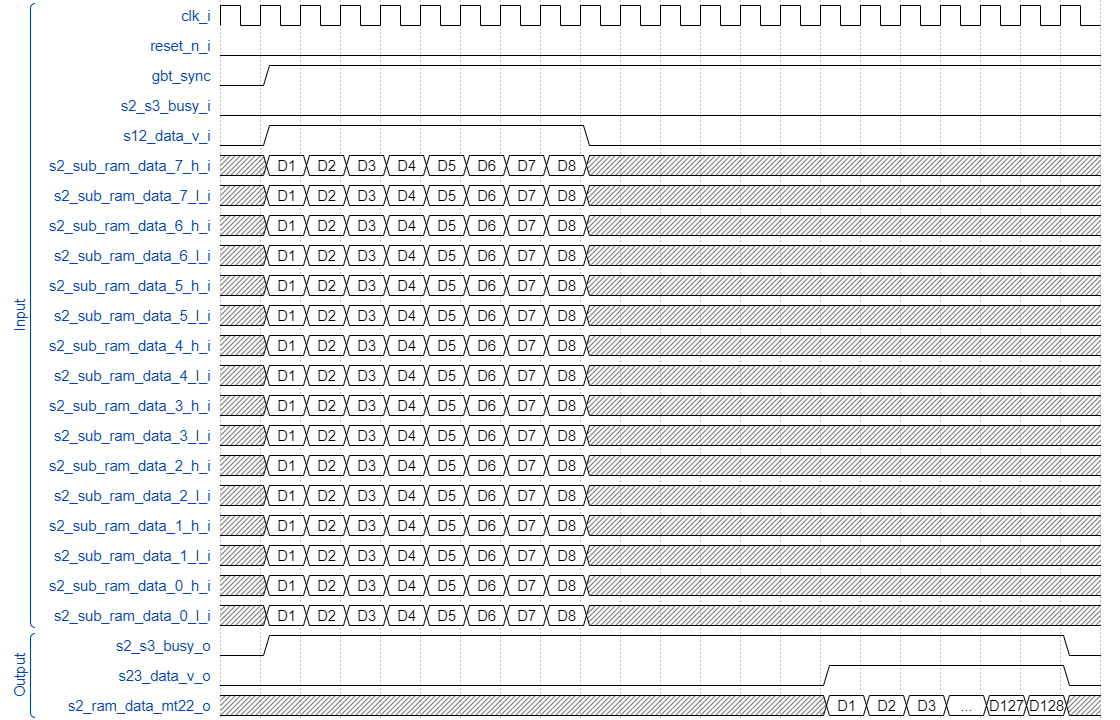}
\caption{Timing diagram depicting test data being driven through Stage 2 readout input ports and output}
\label{fig:s2_wav}
\end{figure}

Evaluation of architecture presented in \ref{sec:data_ro} for Stage 2 readout indicates acceptable conformance to the requirements established. It can be seen in the timing diagram (Fig. \ref{fig:s2_wav}) that successful merging and sorting of data originating from a single GBT link is achieved.

The S2 RAM module is implemented using the M20K internal memory of the Arria 10 FPGA. Using this memory in dual-port mode, accommodating the 16 GBT links and 4 planes results in 32 dual-port RAM modules. As the RAM modules require a width of 64 bits and a depth of 16 words (local card 1 to 16) the memory usage can be calculated as follows:
\begin{equation}
\resizebox{.9\hsize}{!}{$
    \begin{split}
    Memory\ Used & = (32\ RAM\ modules)*(16\ words\ depth)*(64\ bits) \\
    & = \ 0.004096\ MB
    \end{split}
    $}
\end{equation}
Considering that M20K makes up $\sim$7MB of internal memory, the Stage 2 readout module utilizes 0.06\% of available memory.
\newline \newline
The latency of the S2 RAM module is calculated from the time the last dataset is written to the time the last dataset is read.
This occurs at the 8\textsuperscript{th} and 141\textsuperscript{st} clock cycle respectively. The latency is therefore 133 clock cycles or $\sim$ 554ns.

\subsubsection{Stage 3 Readout:}

\begin{figure}[h!]
\centering
\includegraphics[width=0.7\textwidth]{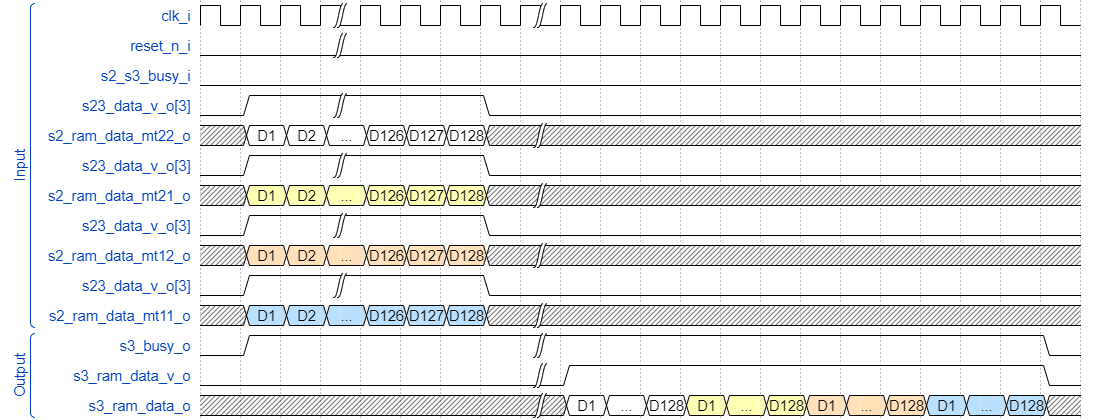}
\caption{Stage 3 Readout merging Stage 2 Readout data from 4 S2-RAM module output busses from 4 planes into a single output bus}
\label{fig:s3_wav}
\end{figure}

Evaluation of architecture presented in \ref{sec:data_ro} for Stage 3 readout indicates acceptable conformance to the requirements established. It can be seen in the timing diagram (Fig. \ref{fig:s3_wav}) that successful merging and sorting of data originating from a single GBT link is achieved.

Stage 3 readout receives the data payload from stage 2 of all 4 planes simultaneously, temporarily writing them to single-port RAM modules. Once this process is complete the RAM modules for each plane are read out consecutively merging the data into one data stream and ordering them from plane MT11 to M22.

Considering the S3 RAM module as a single 4-port RAM, the figure shows the data payload from each column (plane) of S2 module being written to the 4 segments of the S3 RAM in sync. After the anticipated RAM buffering from the first RAM module, the entire S3 memory is read linearly from bottom to top on each clock cycle operating as desired.
\newline \newline
The S3 RAM module similarly to the S2 RAM module was implemented using M20K memory. Using 4 blocks of memory in single-port mode, interfaced to the 4 output ports of the S2 module the memory usage can be calculated as follows:
\begin{equation}
\resizebox{.9\hsize}{!}{$
    \begin{split}
    Memory\ Used &= (4\ RAM\ modules)*(128\ words\ depth)*(64\ bits) \\
    &=\ 0.004096\ MB
    \end{split}
    $}
\end{equation}

Again, noting that M20K makes up $\sim$7MB of internal memory, the Stage 3 readout module utilizes 0.06\% of available memory
\newline \newline
The latency of the S2 RAM module is calculated from the time the last dataset is written to the time the last dataset is read.
This occurs at the 128\textsuperscript{th} and 648\textsuperscript{th} clock cycle, respectively. The latency is therefore 520 clock cycles or $\sim$ 2167ns.

\section{Conclusion and Future Work}

In this section, an analysis of conformance to the requirements is performed, as established previously in this report. Furthermore, future work exists for a complete and functional system which is elucidated later in this section.

\subsection{Conclusions}
\label{sec:conc}
The ALICE MID user logic architecture proposed in this work adequately addresses the challenges associated with high-throughput data steaming. The proposed architecture can effectively extract, reformat, synchronize and readout data transmitted from the MID detector, performing zero suppression and data reduction as required. Although there is still prototyping work needed, the architecture that was developed and tested shows a functionally sufficient design well within device resource constraints.  
\newline \newline
The modularized design employed provides two distinct advantages by leveraging parameterization. Firstly, it can be scaled to test any number of GBT links significantly reducing adaptation time for various testing methodologies and secondly allowing easy adaptation for other detectors with similar requirements by allowing developers easy access to individual components. Additionally, more features can be easily included in the architecture, such as clustering algorithms.

\subsection{Future Work}
\label{sec:fut_wrk}
The prototype developed in this work essentially represents a data routing network with the exception of the data reformatting. As such, future work should focus on architectural development and the development of additional features such as clustering. As this prototype was tailored to the FET trigger, future work should address the other triggers of the system. Additionally, a FET produces a full data set where as in normal operation this is rarely the case and the user logic should perform zero suppression on the payload. Futhermore, population of the RDH was not performed in this work and needs to be implemented including calculation of the HBID. Finally, packetization and transmission of the data payload is required for full system integration.

\end{document}